\documentclass[fleqn,twoside]{article}
\usepackage{espcrc2}
\usepackage{cite}
\usepackage{amsmath}
\usepackage{amssymb}

\readRCS
$Id: espcrc2.tex,v 1.2 2004/02/24 11:22:11 spepping Exp $
\usepackage{graphicx}

\newcommand\SH{\,\mbox{$\sqcup \! \sqcup$}\,}

\newcommand{\Li}{{\rm Li}}
\newcommand{\Mvec}{{\bf \rm M}}

\title{\vspace*{-10mm}
{\footnotesize DESY 08--087; 
SFB/CPP-08-43 
\hfill {\tt arxiv:0807.0700 [math-ph]}
}\\
Structural Relations between Nested Harmonic Sums
       \thanks{Presented at Loops and Legs in
               Quantum Field Theory, Sondershausen, Germany, 2008. This paper 
was supported in part by SFB-TR-9:
                 Computergest\"utze Theoretische Teilchenphysik.
}%
        }
\author{Johannes Bl\"umlein\address[DESY]{Deutsches Elektronen-Synchrotron, 
DESY,
                     Platanenallee 6, D-15738 Zeuthen, Germany},
}

\runtitle{  }
\runauthor{ }
\begin{document}
\begin{abstract}
\noindent  
We describe the structural relations between nested harmonic sums emerging
in the description of physical single scale quantities up to the 3--loop
level in renormalizable gauge field theories. These are weight {\sf w=6}
harmonic sums. We identify universal basic functions which allow to describe
a large class of physical quantities and derive their complex analysis.
For the 3--loop QCD Wilson coefficients 35 basic functions are required, 
whereas a subset of 15 describes the 3--loop anomalous dimensions.
\vspace{1pc}
\end{abstract}
%
\maketitle
%
%
\section{Introduction}

\noindent
Scattering cross sections in renormalizable Quantum Field Theories which depend on 
a single kinematic or mass scale obey a particularly simple form if dealt with in Mellin 
space. Contrary to the case in momentum-fraction space, where the variable 
$z = p/P$
is referred to, or a representation depending on $z= s/t$, with $s$ and $t$ Mandelstam 
variables, the results are expressed in terms of nested harmonic sums with an outer
summation index $N$~\cite{HSUM}. In this way a unique language is available in which the 
integrals emerging in higher order calculations can be expressed. Although it is
not expected that even in this case the representation will be sufficient, cf. 
\cite{JB08}, it holds for all 3-loop calculations having been performed so 
far \cite{THR1}. Furthermore, this representation holds independent of the processes
considered.

The nested harmonic sums are recursively defined by
\begin{eqnarray}
S_{b,\vec{a}}(N) = \sum_{k=1}^N \frac{{\rm sign}(b)^k}{k^{|b|}} S_{\vec{a}}(k)~,
\end{eqnarray}
where $\vec{a} = (a_1,...,a_l)$ and $b,a_i~\in~\mathbb{Z} \backslash \{0\}~.$
The weight {\sf w} and depth {\sf d} of a harmonic sum $S_{\vec{a}}(N)$ are given by
${\sf w} = \sum_{k=1}^l |a_k|$ and ${\sf d} = l$. At each weight {\sf w} there are
$2 \cdot 3^{\sf w-1}$ harmonic sums. In the limit $N \rightarrow \infty$ the nested harmonic 
sums turn into multiple zeta values $\zeta_{\vec{a}}$,~\cite{ZETA}. 
As well known, there is 
a large number of algebraic relations between the multiple zeta values due to the shuffle- 
and stuffle relations, cf.~\cite{BBBL}. The nested harmonic sums form a quasi--shuffle 
algebra,~\cite{HOF}, with the shuffle product $\SH$, which maps two 
index sets in all possible orders preserving the order of the original sets. The algebraic 
relations of the harmonic sums are well known, \cite{ALG}, and are 
related to the shuffle product and the structure of the index sets. The number 
of the respective
basis elements are counted by the Lyndon words or due to the Witt relations \cite{WITT}.
Let us consider the harmonic sums up to weight {\sf w=6}. The reduction 
obtained cumulatively by 
the algebraic relations $\#_r$  compared to the original number $\#_c$ is
\begin{center}
\begin{tabular}{r|r|r|r|r|r|r}
w    &  1 &  2  &  3 &  4 &   5 &   6 \\
\hline
$\#_c$ &  2 &  8  & 26 & 80 & 242 & 728 \\
$\#_r$ &  2 &  5  & 13 & 31 &  79 & 195
\end{tabular} 
\end{center}
In the physical applications we observe \cite{DITT,BK1,JB08} that one may find 
representations 
in which the index $\{-1\}$ never appears in a nested harmonic sum. The number of harmonic 
sums
of this type $N_{\neg\{-1\}}(w)$ and the corresponding number of basis elements $N^{\rm 
basis}_{\neg\{-1\}}(w)$ are given by
\begin{eqnarray}
N_{\neg\{-1\}}(w) &=& \frac{1}{2} \left[\left(1-\sqrt{2}\right)^w + 
\left(1+\sqrt{2}\right)^w\right] \nonumber\\
N_{\neg\{-1\}}^{\rm basis}(w) &=& \frac{2}{w} \sum_{d|w} 
\mu\left(\frac{w}{d}\right) N_{\neg\{-1\}}(d)~, 
\end{eqnarray}
with $\mu(d)$ the M\"obius function. This leads to the following reduction
\begin{center}
\begin{tabular}{r|r|r|r|r|r|r}
w    &  1 &  2  &  3 &  4 &   5 &   6 \\
\hline
$\#_c$ &  1 &  4  & 11 & 28 &  69 & 168 \\
$\#_r$ &  1 &  3  &  7 & 14 &  30 &  60
\end{tabular}
\end{center}
Note that already the numbers at input are lower than the numbers after the algebraic 
reduction in the former case.  

All further relations between the nested harmonic sums are called {\sf structural 
relations}. They result from the mathematical structure of these objects beyond that
given by their indices. 
%
%
\section{Structural Relations}

\noindent
In the physical application nested harmonic sums emerge as a consequence of
the light-cone expansion \cite{LC} and related techniques, observing the 
crossing relations for the respective processes, cf. e.g. \cite{BLKO1}. The 
physical quantities are thus defined at either even or odd integers $N$. To 
use these expressions in experimental data analyzes it is, however, necessary
to map these representations back to momentum-fraction space. This requires
an analytic continuation of the nested harmonic sums from the even, resp. odd,
values of $N$ to $N~\in~\mathbb{C}$. Eventually one has to derive the complex 
analysis for the nested harmonic sums. In passing, $N$ takes values 
$N~\in~\mathbb{Q}$ and
$N~\in~\mathbb{R}$, which leads to new relations, as we will outline below.
From a practical point of view, the most complicated part consists in deriving
the corresponding representations for $N~\in~\mathbb{C}$ for a large number of 
harmonic sums. Whenever possible, we will seek equivalence classes for these
sums, the elements of which can be easily accessed applying some operator, which
can be straightforwardly realized even in the final numerical precision 
representations. 

Harmonic sums can be represented in terms of Mellin integrals, cf. \cite{HSUM},
\begin{eqnarray}
S_{\vec{a}}(N) = {\bf \rm  M}[f_{\vec{a}}(x)](N) = \int_0^1 dx 
\frac{\hat{f}_{\vec{a'}}(x)}{1 \pm x}|_{\rm reg} 
x^{N-1}~. \nonumber
\end{eqnarray}
The first structural relation one obtains at {\sf w=1} by decomposing
\begin{eqnarray}
\label{lab1}
\frac{1}{1-x^2} = \frac{1}{2}\left[\frac{1}{1-x} + \frac{1}{1+x}\right]
\end{eqnarray}
which yields
\begin{eqnarray}
\hspace*{-5mm} -\psi\left(\frac{N}{2}\right) &=& - \psi(N) + \beta(N) +\ln(2) \nonumber\\
\hspace*{-5mm} \beta(N) &=& 
\frac{1}{2}\left[\psi\left(\frac{N+1}{2}\right)-\psi\left(\frac{N}{2}\right)\right]~.
\end{eqnarray}
Since $S_{-1}(N) = (-1)^N \beta(N+1) - \ln(2), S_1(N)=\psi(N+1) +\gamma_E$, the sum
$S_{-1}(N)$ is not algebraically independent of $S_1(N)$ for $N~\in~\mathbb{Q}$.
At higher weights (\ref{lab1}) can be generalized using
\begin{eqnarray}
\label{lab2a}
\ln(1-x^2) &=& \ln(1-x) + \ln(1+x) \\
\label{lab2}
\frac{1}{2^{k-2}} \Li_k(x^2) &=& \Li_k(x) + \Li_k(-x)
\end{eqnarray}
as numerator function.

Considering $N~\in~\mathbb{R}$ one may define
the differentiation of harmonic sums via
\begin{eqnarray}
\label{lab3}
\frac{d}{dN} S_{\vec{a}}(N) = {\bf \rm M}[\ln(x) f_{\vec{a}}(x)](N)~.
\end{eqnarray}
In this way we obtain 
\begin{eqnarray}
S_2(N) = - \frac{d}{dN} S_{1}(N)  +\zeta_2~.
\end{eqnarray}
In general one observes that the set of harmonic sums, extended by the multiple zeta
values, is closed under differentiation. In particular one may form 
equivalence classes
through differentiation and consider the harmonic sum with the lowest weight of this class
as its representative. A consequence of these two relations is, that all single harmonic
sums may be traced back to $S_1(N)$. The complex analysis of the whole class derives from
that of $S_1(N)$. 

A third class of structural relations emerges through the respective iterated integral
representation \cite{ITINT}, combined with algebraic relations. At {\sf w=2} one obtains
\begin{align}
\label{lab4}
&\Mvec\left[\frac{\ln(1-x)}{1+x}\right](N) = 
-\Mvec\left[\frac{\ln(1+x)}{1+x}\right](N) 
\nonumber\\
&- [\psi(N) + \gamma_E +\ln(2)]\beta(N) +\beta'(N) 
\end{align}
In this way we identify the first basic function
\begin{eqnarray}
F_1(x) = \frac{\ln(1+x)}{1+x}~,
\end{eqnarray}
which is related to the harmonic sum $S_{-1,1}(N)$. Although in physics problems
the index $\{-1\}$ is not emerging,  $F_1(x)$ is still useful to express a 
series
of harmonic sums. All {\sf w=2} harmonic sums reduce algebraically or are related
to $F_1(x)$ and single harmonic sums.

At {\sf w=3} we apply (\ref{lab2}) for the first time. Together with a relation similar to
(\ref{lab4}) all sums which do not contain $\{-1\}$ as index reduce to the Mellin transforms
of
\begin{eqnarray}
F_{2,3}(x) = \frac{\Li_2(x)}{1 \pm x}~.
\end{eqnarray}

Let us now study the double sums in general. Using the above relations one may
show that Nielsen integrals \cite{NIELS} are sufficient to express all double 
sums in terms of Mellin transformations of the functions
\begin{eqnarray}
\hat{F}_{k,\pm}(x) = \frac{\Li_k(x)}{1 \pm x}~.
\end{eqnarray}
For even weight {\sf w} there is an Euler relation and only $\Li_k(x)/(1+x)$ contributes.
Let us illustrate this for $S_{2,3}(N)$. One derives
\hspace*{-5mm}{\small
\begin{eqnarray}
\lefteqn{S_{2,3}(N) = 
3 \zeta_4 S_1(N)~~+} \nonumber\\
&& \hspace*{-7mm} \Mvec\left[\left[\frac{\ln(x)\left[S_{1,2}(1-x) \zeta_3\right]
+3 [S_{1,3}(1-x) - \zeta_4]}{x-1}\right]_+\right] \hspace*{-1mm}(N) \nonumber
\end{eqnarray}}
\normalsize

\noindent
where the functions $S_{1,k}(1-x)$ are polynomials of $\ln(x), \ln(1-x)$ and $\Li_k(x)$. 
Therefore
$S_{2,3}(N)$ is related to $S_{4,1}(N)$ up to derivatives of known harmonic 
sums of lower
weight.

In the following we only have to consider sums with depth {\sf d $\geq$ 3}. 
We derive all one-dimensional integral representations for the harmonic sums
up to {\sf w=6} which do not contain an index $\{-1\}$ and use the above relations. 
We obtain the following set of {\sf basic functions}, \cite{JB08,JB08a}.

{\small \begin{alignat}{2}
{\sf w=1:}~~~& 1/(x-1)_+                                                         
\nonumber\\
{\sf w=2:}~~~ & \ln(1+x)/(x+1)                                                    
\nonumber\\
{\sf w=3:}~~~ & \Li_2(x)/(x \pm 1)                                                
\nonumber\\
{\sf w=4:}~~~ & \Li_3(x)/(x + 1),    
\nonumber\\
& S_{1,2}(x)/(x \pm 1)                         
\nonumber
\\
{\sf w=5:}~~~ & \Li_4(x)/(x \pm 1),   \nonumber\\
& S_{1,3}(x)/(x \pm 1)  \nonumber\\                       
              & S_{2,2}(x)/(x \pm 1), \nonumber\\   
& \Li_2^2(x)/(x \pm 1)                      
\nonumber
\end{alignat}}{\small \begin{alignat}{2}
\label{eq:bas}
              & [\ln(x) S_{1,2}(-x) - \Li^2_2(-x)/2]/(x \pm 1)                         
\nonumber\\
{\sf w=6:}~~~ & \Li_5(x)/(x + 1), \nonumber\\
& S_{1,4}(x)/(x \pm 1)  \nonumber\\                       
              & S_{2,3}(x)/(x \pm 1),   \nonumber\\
& S_{3,2}(x)/(x \pm 1) \nonumber\\                        
              & \Li_2(x) \Li_3(x)/(x \pm 1)\nonumber\\                         
              & S_{1,2}(x)\Li_2(x)/(x \pm 1)\nonumber\\                         
&A_1(x)/(x+1) \nonumber\\
&A_2(x)/(x \pm 1) \nonumber\\
&A_3(x)/(x+1) \nonumber\\
&H_{0,-1,0,1,1}(x)/(x-1) \nonumber\\
&[A_1(-x)+N_{\alpha}(x)]/(x+1)~,
\nonumber\\
\end{alignat}} \normalsize
where
\begin{eqnarray}
A_1(x) &=& \int_0^x \frac{dy}{y} \Li_2^2(y) \nonumber\\
A_2(x) &=& \int_0^x \frac{dy}{y} \ln(1-y) S_{1,2}(y) \nonumber\\
A_3(x) &=& \int_0^x \frac{dy}{y} [\Li_4(y) -\zeta_4]~,
\end{eqnarray}
and $\left. N_{\alpha}(x)\right|_{\alpha=1...3}$ are polynomials of Nielsen integrals. 
Up to {\sf w=5} the numerator functions are Nielsen integrals, while at {\sf w=6}
some of the numerator functions are general harmonic polylogarithms over the alphabet
$\{0,1,-1\}$ \cite{VR} in the representation we derived. Yet it may still be, that an 
equivalent
representation can be found over a two-letter alphabet, which will be investigated 
further.~\footnote{Structures of this type have been found at least in case of the multiple 
zeta 
values recently \cite{BBV}.}

The Mellin transforms of the {\sf basic functions} in (\ref{eq:bas}) span the space
of the harmonic sums which occur in higher order calculations for single scale 
quantities in Quantum Chromodynamics. Examples are the anomalous dimensions and Wilson
coefficients in deeply-inelastic scattering up to 3-loops~\cite{THR1,THR2}.
With rising order in the coupling constant $\alpha_s$ the number of contributing basic 
functions is given by
\begin{alignat}{2}
O(\alpha_s)   &~~ {\rm Wilson~coeff./anom.~dim.} & {\#~~~~~1}
\nonumber\\
O(\alpha_s^2) &~~ {\rm anomalous~dimensions}     & {\#~~~~~2}
\nonumber
\end{alignat} 
\begin{alignat}{2}
O(\alpha_s^2) &~~ {\rm Wilson~coefficients }     & {\#~\leq 5}
\nonumber\\
O(\alpha_s^3) &~~ {\rm anomalous~dimensions}     & {\#~~~~15}
\nonumber\\
O(\alpha_s^3) &~~ {\rm Wilson~coefficients }     & {\#~~~~35}\nonumber\\
\end{alignat}
We represented a wide class of massless and massive 2-loop quantities in terms of harmonic 
sums, among them the unpolarized and polarized Drell-Yan cross section and the hard 
scattering cross sections for hadronic Higgs-boson and pseudoscalar-boson production in the
heavy top quark limit \cite{DYHIG}, the unpolarized and polarized time-like anomalous 
dimensions and Wilson coefficients \cite{TIME}, the polarized anomalous dimensions and Wilson 
coefficients \cite{BM}, the heavy flavor deep-inelastic Wilson coefficients in the limit 
$Q^2 \gg m^2$ \cite{HEAV}, including the $O(\varepsilon)$ terms, as well as the virtual- and 
soft corrections to Bhabha scattering \cite{BK1}. All these quantities fall into the 
third class above and
are represented by the respective basic functions and simple polynomial factors in the 
Mellin variable.
The same structure is obtained for other 
similar processes more. 
%
%
\section{Complex Analysis}

\noindent
Finally we have to derive the analytic continuation of the nested harmonic sums
$N \in \mathbb{N} \rightarrow N \in \mathbb{C}$. In the past precise numerical 
representations were derived in \cite{ANCONT}.~\footnote{This is also possible in the case 
of the heavy quark Wilson coefficients, including power corrections, cf.~\cite{AB}.}
Here the integrand is dealt with using the {\tt MINIMAX}-method 
\cite{MINIMAX}, which leads
to an adaptive representation in the complex plane in terms of a rational function.
We seek, however, for a rigorous representation.

It is well-known \cite{FACSE} that Mellin transforms of the type
\begin{eqnarray}
\Omega(z) &=& \int_0^1~dt~t^{z-1}~\varphi(t);\nonumber\\ 
\varphi(1-t) &=& \sum_{k=0}^\infty a_k t^k~, 
\end{eqnarray}
are factorial series,
\begin{eqnarray}
\Omega(z) = \sum_{k=0}^\infty \frac{a_{k+1} k!}{z(z+1) \ldots (z+k)}~.
\end{eqnarray}
$\varphi(t)$ has to be analytic at $t=1$. $\Omega(z)$ is a meromorphic function for 
$z~\in~\mathbb{C}$, with poles at the non-positive integers. It obeys a recursion for
$z \rightarrow z+1$ and has an analytic asymptotic representation. The choice of the basic 
functions in (\ref{eq:bas}) is not immediately suited in all cases, as some of the functions
possess branch points at $t=1$. However, one may map this behaviour using relations for 
the basic functions, as for the change of the argument $t \rightarrow (1-t)$. 
Let us consider the example 
\begin{eqnarray}
F_3(N) =\Mvec[\Li_2(x)/(1+x)](N)~.
\end{eqnarray}
The recursion relation is given by
\begin{eqnarray}
\hspace*{-0.7cm}
F_3(z+1) &=& - F_3(z) 
\nonumber\\ & &
+ \frac{1}{z}\left[\zeta_2 - \frac{\psi(z+1) + 
\gamma_E}{z}\right]~.
\end{eqnarray}
We map 
\begin{eqnarray}
\Li_2(z) \rightarrow - \Li_2(1-z) - \ln(z) \ln(1-z) +\zeta_2~,
\end{eqnarray}
and derive the asymptotic representation for
\begin{eqnarray}
\Mvec\left[\frac{\Li_2(1-z)}{1+z}\right](N) &\sim& \frac{1}{2N^2} + \frac{1}{4N^3}- 
\frac{7}{24}\frac{1}{N^4}\nonumber\\ &&
-\frac{1}{3}\frac{1}{N^5} + \frac{73}{120} \frac{1}{N^6} 
\ldots \nonumber\\
\end{eqnarray}
We still need to consider $\Mvec[\ln(z)\ln(1-z)/(1+z)](N)$, which is a derivative of 
the function $F_1(N)$ and a few simpler terms. Representations like this can be obtained
for all the {\sf basic functions}. In this way analytic expressions for the Mellin 
transforms, resp. nested harmonic sums, which are needed up to the level {\sf 
w=6}, are obtained.
The above relations can be tuned to any numerical accuracy by analytic means. 
%
%
\section{Conclusions}

\noindent
In perturbative higher order calculations in renormalizable Quantum Field Theories
nested harmonic sums form an appropriate way of representation for single-scale quantities, 
at least up to 3--loop orders. The number of multiple nested harmonic sums grows 
exponentially with the
weight {\sf w}. The harmonic sums obey algebraic and structural relations, which 
reduce the number of objects to a small set of {\sf basic functions}. The 
quantities
are meromorphic functions in $\mathbb{C}$ with poles at the non-positive integers, obey
recursion relations and possess an analytic asymptotic representation. Due to this these
functions are known in analytic form. The basic functions emerge as unique quantities
in a large variety of higher order calculations and hint to a rather simple 
mathematical structure behind Feynman diagrams. The simple pattern, known in 
case of 
zero-scale quantities, where only a few basis elements 
span the space of all multiple zeta values,
is rediscovered for the next more complicated class of processes and found 
to be of quite similar structure. It is an observation that harmonic sums with an index $\{-1\}$
can be avoided in the physical quantities, for yet unknown reason.
Due to this the complexity of 728 objects at {\sf w=6}, anticipated originally, reduces to 
168. The latter ones can now be represented by 35 basic functions, which is an essential 
compactification. 


\end{document}